\documentclass[11pt]{article}
\usepackage{latexsym}
\newcommand{\eproof}{\mbox{\ }\hfill $\Box$ \par \vskip 10pt}

\newtheorem{Theorem}{Theorem}[section]
\newtheorem{lemma}[Theorem]{Lemma}
\newtheorem{prop}[Theorem]{Proposition}

\begin{document}

\title{Low frequency dispersive estimates for the Schr\"odinger 
group in higher dimensions}

\author{{\sc Simon Moulin and Georgi Vodev}}

\date{}

\maketitle

\abstract{For a large class of real-valued potentials, $V(x)$, $x\in
{\bf R}^n$, $n\ge 4$ , we prove dispersive estimates for the low
frequency part of $e^{it(-\Delta+V)}P_{ac}$, provided the zero is
neither an eigenvalue nor a resonance of $-\Delta+V$, where $P_{ac}$
is the spectral projection onto the absolutely continuous spectrum of
$-\Delta+V$. This class includes potentials $V\in L^\infty({\bf
R}^n)$ satisfying $V(x)=O\left(\langle
x\rangle^{-(n+2)/2-\epsilon}\right)$, $\epsilon>0$. As a
consequence, we extend the results in \cite{kn:JSS} to a larger
class of potentials.}

\setcounter{section}{0}
\section{Introduction and statement of results}

Let $V\in L^\infty({\bf R}^n)$, $n\ge 4$, be a real-valued function
satisfying $$|V(x)|\le C\langle x\rangle^{-\delta},\quad\forall x\in
{\bf R}^n,\eqno{(1.1)}$$ with constants $C>0$, $\delta>(n+2)/2$.
Denote by $G_0$ and $G$ the self-adjoint realizations of the
operators $-\Delta$ and $-\Delta+V$ on $L^2({\bf R}^n)$,
respectively. It is well known that the absolutely continuous
spectrums of the operators $G_0$ and $G$ coincide with the interval
$[0,+\infty)$, and that $G$ has no embedded strictly positive
eigenvalues nor strictly positive resonances. However, $G$ may have
in general a finite number of non-positive eigenvalues and that the
zero may be a resonance. We will say that the zero is a regular
point for $G$ if it is neither an eigenvalue nor a resonance in the
sense that the operator $1-V\Delta^{-1}$ is invertible on $L^1$ with
a bounded inverse. Let $P_{ac}$ denote the spectral projection onto the
absolutely continuous spectrum of $G$. When $n\ge 3$,  Journ\'e,
Sofer and  Sogge \cite{kn:JSS} proved the following dispersive
estimate
$$\left\|e^{itG}P_{ac}\right\|_{L^1\to L^\infty}\le
C|t|^{-n/2},\quad t\neq 0,\eqno{(1.2)}$$ provided the zero is
neither an eigenvalue nor a resonance, for potentials satisfying
(1.1) with $\delta>n+4$ as well as the condition $$\widehat V\in
L^1.\eqno{(1.3)}$$ This was later improved by Yajima \cite{kn:Y1}
for potentials satisfying (1.1) with $\delta>n+2$. When $n=3$, the
estimate (1.2) in fact holds without (1.3). In this case, it was
proved in \cite{kn:GS} for potentials satisfying (1.1) with
$\delta>3$ and was later improved in \cite{kn:V1} and \cite{kn:Y2}
for potentials satisfying (1.1) with $\delta>5/2$. Goldberg
\cite{kn:G} has recently showed that (1.2) holds for potentials
$V\in L^{3/2-\epsilon}\cap L^{3/2+\epsilon}$, $0<\epsilon\ll 1$,
which includes potentials satisfying (1.1) with $\delta>2$. When
$n=2$, (1.2) is proved by Schlag \cite{kn:S} for potentials
satisfying (1.1) with $\delta>3$.

Given any $a>0$, set $\chi_a(\sigma)=\chi_1(\sigma/a)$, where
$\chi_1\in C^\infty({\bf R})$, $\chi_1(\sigma)=0$ for $\sigma\le 1$,
$\chi_1(\sigma)=1$ for $\sigma\ge 2$. Set $\eta_a=\chi(1-\chi_a)$,
where $\chi$ denotes the characteristic function of the interval
$[0,+\infty)$. Clearly, $\eta_a(G)+\chi_a(G)=P_{ac}$. When $n\ge 4$,
 dispersive estimates with loss of $(n-3)/2$ derivatives for the
operator $e^{itG}\chi_a(G)$, $\forall a>0$, have been recently
proved in \cite{kn:V2} under the assumption (1.1), only. The loss of
derivatives in this case is a high frequency phenomenon and cannot
be avoided unless one imposes some regularity condition on the
potential (see \cite{kn:GV}). The condition (1.3) in \cite{kn:JSS}
plays this role but it seems too strong. The natural conjecture
would be that we have dispersive estimates for $e^{itG}\chi_a(G)$
with loss of $\nu$ derivatives, $0\le\nu\le (n-3)/2$, provided $V\in
C^{(n-3)/2-\nu}({\bf R}^n)$ (with a suitable decay at infinity). It
turns out that no regularity on the potential is needed in order to
get dispersive estimates for the low frequency part
$e^{itG}\eta_a(G)$, $a>0$ small. One just needs some decay at
infinity. In fact, the low frequency analysis turns out to be easier
in dimensions $n\ge 4$ compared with the cases of $n=2$ and $n=3$,
and can be carried out for a larger class of potentials satisfying
(with some $0<\epsilon\ll 1$)
$$\sup_{y\in {\bf R}^n}\int_{{\bf
R}^n}\left(|x-y|^{-n+2}+|x-y|^{-(n-2)/2+\epsilon}\right)|V(x)|dx\le
C<+\infty.\eqno{(1.4)}$$ Clearly, (1.4) is fulfilled for potentials
satisfying (1.1). Our main result is the following

\begin{Theorem} Let $n\ge 4$, let $V$ satisfy (1.4) and assume that the
 zero is a
regular point for $G$. Then, there exists a constant $a_0>0$ so that
for $0<a\le a_0$ we have the estimate
$$\left\|e^{itG}\eta_a(G)\right\|_{L^1\to L^\infty}\le
C|t|^{-n/2},\quad t\neq 0.\eqno{(1.5)}$$
\end{Theorem}

\noindent
 {\bf Remark 1.} We expect that (1.5) holds true for the
larger class of potentials satisfying
$$\sup_{y\in {\bf R}^n}\int_{{\bf
R}^n}\left(|x-y|^{-n+2}+|x-y|^{-(n-1)/2}\right)|V(x)|dx\le
C<+\infty,\eqno{(1.6)}$$
 but the proof in this case would require a
different approach.

Combining (1.5) with the estimates of \cite{kn:V2}, we obtain the
following

\begin{Theorem}  Let $n\ge 4$, let $V$ satisfy (1.1) and assume that 
the zero is a
regular point for $G$. Then, we have the estimates, $\forall t\neq
0$, $0<\epsilon\ll 1$,
$$\left\|e^{itG}P_{ac}f\right\|_{L^\infty}\le
C|t|^{-n/2}\left\|\langle
G\rangle^{(n-3)/4}f\right\|_{L^1},\eqno{(1.7)}$$
$$\left\|e^{itG}P_{ac}f\right\|_{L^\infty}\le
C_\epsilon|t|^{-n/2}\left\|\langle
x\rangle^{n/2+\epsilon}f\right\|_{L^2}.\eqno{(1.8)}$$
\end{Theorem}

\noindent
 {\bf Remark 2.} The proof in \cite{kn:V2} is based on
uniform estimates for the operator $\psi(h^2G)$, $0<h\le 1$,
$\psi\in C_0^\infty((0,+\infty))$ (see Lemma 2.2 of \cite{kn:V2} or
Lemma 2.3 of \cite{kn:V3}). In the proof of this lemma (which is
given in \cite{kn:V3}), however, there is a mistake. That is why, we
will give a new proof in Appendix 1 of the present paper.

\noindent
 {\bf Remark 3.} We conjecture that the estimates (1.7) and (1.8)   
 hold true for potentials satisfying (1.1) with $\delta>(n+1)/2$.

Theorem 1.1 also allows to extend the results in \cite{kn:JSS} to a
larger class of potentials. More precisely, we have the following

\begin{Theorem}  Let $n\ge 4$, let $V$ satisfy (1.1) with $\delta>n-1$
as well as (1.3), and assume that the zero is a regular point for
$G$. Then, the estimate (1.2) holds true.
\end{Theorem}

Theorem 1.3 follows from (1.5) and the dispersive estimate for
$e^{itG}\chi_a(G)$ proved in Appendix 2.

To prove (1.5) we adapt the {\it semi-classical} approach of
\cite{kn:V2} based on the {\it semi-classical} version of Duhamel's
formula (which in our case is of the form (3.4) or (3.5)). While in
\cite{kn:V2} the estimates had to be uniform with respect to the
semi-classical parameter $0<h\le 1$, in the case of low frequency we
need to make them uniform for $h\gg 1$ (see (3.1)). This, however,
turns out to be easier (when $n\ge 4$) as we can absorb the
remaining terms taking $h$ big enough (see Section 3). That is why,
we do not need any more to work on weighted $L^2$ spaces (as in
\cite{kn:V2}), which in turn allows to cover a much larger class of
potentials. As mentioned in Remark 1, the natural class of
potentials for which the low frequency analysis works out (for $n\ge
4$) is given by (1.6), and the fact that the crucial Proposition 2.1
below holds true under (1.6) is a strong indication for that. In
fact, (1.4) is used in the proof of Proposition 2.3, only.

\section{Preliminary estimates}

Let $\psi\in C_0^\infty((0,+\infty))$. We will first prove the
following

\begin{prop}
Let $n\ge 4$, let $V$ satisfy (1.6) and assume that the zero is a
regular point for $G$. Then, there exist positive constants
$C,\beta$ and $h_0$ so that the following estimates hold
$$\left\|\psi(h^2G_0)\right\|_{L^1\to L^1}\le C,\quad
h>0,\eqno{(2.1)}$$
$$\left\|\psi(h^2G)\right\|_{L^1\to L^1}\le C,\quad
h\ge h_0,\eqno{(2.2)}$$
$$\left\|\psi(h^2G)-\psi(h^2G_0)T\right\|_{L^1\to L^1}\le Ch^{-\beta},\quad
h\ge h_0,\eqno{(2.3)}$$ where the operator
$$T=\left(1-V\Delta^{-1}\right)^{-1}:L^1\to L^1\eqno{(2.4)}$$
is bounded by assumption.
\end{prop}

{\it Proof.} Set $\varphi(\lambda)=\psi(\lambda^2)$. We are going to
take advantage of the formula
$$\psi(h^2G)=\frac{2}{\pi}\int_{{\bf
C}}\frac{\partial\widetilde\varphi}{\partial\bar
z}(z)(h^2G-z^2)^{-1}zL(dz),\eqno{(2.5)}$$ where $L(dz)$ denotes the
Lebesgue measure on ${\bf C}$, $\widetilde\varphi\in C_0^\infty({\bf
C})$ is an almost analytic continuation of $\varphi$ supported in a
small complex neighbourhood of supp$\,\varphi$ and satisfying
$$\left|\frac{\partial\widetilde\varphi}{\partial\bar
z}(z) \right|\le C_N|{\rm Im}\,z|^N,\quad\forall N\ge 1.$$ For
$\pm{\rm Im}\,z>0$, denote
$${\cal R}_{0,h}^\pm(z)=(h^2G_0-z^2)^{-1},\quad {\cal
R}_{h}^\pm(z)=(h^2G-z^2)^{-1}.$$ The kernel of the operator ${\cal
R}_{0,h}^\pm(z)$ is of the form $R_h^\pm(|x-y|,z)$, where
$$R_h^\pm(\sigma,z)=\pm
h^{-2}\frac{i\sigma^{-2\nu}}{4(2\pi)^\nu}{\cal H}_\nu^\pm(\sigma
z/h)=h^{-n}R_1^\pm(\sigma h^{-1},z),$$ where $\nu=(n-2)/2$, ${\cal
H}_\nu^\pm(\lambda)=\lambda^\nu H_\nu^\pm(\lambda)$, $H_\nu^\pm$
being the outgoing and incoming Henkel functions of order $\nu$. It
is well known that these functions satisfy the bound
$$\left|{\cal H}_\nu^\pm(\lambda)\right|
\le C\langle \lambda\rangle^{(n-3)/2}e^{-|{\rm
Im}\,\lambda|},\quad\forall\lambda,\,\pm{\rm Im}\,\lambda\ge
0,\eqno{(2.6)}$$ while near $\lambda=0$ they are of the form
$${\cal H}_\nu^\pm(\lambda)=
a_{\nu,1}^\pm(\lambda)+\lambda^{n-2}\log\lambda\,
a^\pm_{\nu,2}(\lambda),\eqno{(2.7)}$$ where $a_{\nu,j}^\pm$ 
are analytic functions, $a_{\nu,2}^\pm\equiv 0$ if $n$ is odd. By (2.6)
and (2.7), we have
$$\left|{\cal H}_\nu^\pm(\lambda)-{\cal H}_\nu^\pm(0)\right|
\le C|\lambda|^{1/2}\langle
\lambda\rangle^{(n-4)/2},\quad\forall\lambda,\,\pm{\rm
Im}\,\lambda\ge 0.\eqno{(2.8)}$$ Hence, the functions $R_h^\pm$
satisfy the bounds (for $z\in {\bf
C}^\pm_\varphi:=\{z\in{\rm supp}\,\widetilde\varphi,\,\pm{\rm Im}\,z\ge 0\}$,
 $\sigma>0$, $h\ge 1$)
$$\left|R^\pm_h(\sigma,z)\right|\le
Ch^{-2}\left(\sigma^{-n+2}+\sigma^{-(n-1)/2}\right),\eqno{(2.9)}$$
$$\left|R^\pm_h(\sigma,z)-R^\pm_h(\sigma,0)\right|\le
Ch^{-5/2}\left(\sigma^{-n+5/2}+\sigma^{-(n-1)/2}\right).\eqno{(2.10)}$$
 Using the above bounds we will prove the following

\begin{lemma} For $z\in{\bf C}^\pm_\varphi$, we have
$$\left\|V{\cal R}_{0,h}(z)\right\|_{L^1\to L^1}\le Ch^{-2},
\quad h\ge 1,\eqno{(2.11)}$$
$$\left\|V{\cal R}_{0,h}(z)-V{\cal R}_{0,h}(0)
\right\|_{L^1\to L^1}\le Ch^{-5/2},\quad h\ge 1,\eqno{(2.12)}$$
$$\left\|V{\cal R}_{h}(z)\right\|_{L^1\to L^1}\le Ch^{-2},
\quad h\ge h_0,\eqno{(2.13)}$$
$$\left\|{\cal
R}^\pm_{0,h}(z)\right\|_{L^1\to L^1}\le C|{\rm Im}\,z|^{-q},\quad
h>0,\,{\rm Im}\,z\neq 0,\eqno{(2.14)}$$
$$\left\|{\cal
R}^\pm_{h}(z)\right\|_{L^1\to L^1}\le C|{\rm Im}\,z|^{-q},\quad h\ge
h_0,\,{\rm Im}\,z\neq 0,\eqno{(2.15)}$$with constants $C,q,h_0>0$
independent of $z$ and $h$.
\end{lemma}

{\it Proof.} In view of (2.9), the norm in the LHS of (2.11) is
upper bounded by
$$\sup_{y\in{\bf R}^n}\int_{{\bf R}^n}|V(x)|\left|R^\pm_h(|x-y|,z)
\right|dx$$ $$\le Ch^{-2} \sup_{y\in{\bf R}^n}\int_{{\bf
R}^n}|V(x)|\left(|x-y|^{-n+2}+|x-y|^{-(n-1)/2}\right)dx\le
Ch^{-2}.$$ The estimate (2.12) follows in the same way using (2.10).
To prove (2.14), we use (2.6) to get (for $z\in {\bf
C}^\pm_\varphi$, ${\rm Im}\,z\neq 0$)
$$\left|R_1^\pm(\sigma,z)\right|\le C\sigma^{-2\nu}\langle
\sigma\rangle^{(n-3)/2}e^{-\sigma|{\rm Im}\,z|}\le
C\sigma^{-2\nu}\langle \sigma\rangle^{-5/2}|{\rm
Im}\,z|^{-(n+2)/2}.\eqno{(2.16)}$$ By (2.16), the norm in the LHS of
(2.14) is upper bounded by
$$C\int_0^\infty\sigma^{n-1}\left|R_h^\pm(\sigma,z)\right|d\sigma=
C\int_0^\infty\sigma^{n-1}\left|R_1^\pm(\sigma,z)\right|d\sigma$$
 $$\le C|{\rm Im}\,z|^{-(n+2)/2}\int_0^\infty\langle
\sigma\rangle^{-3/2}d\sigma\le C|{\rm Im}\,z|^{-(n+2)/2}.$$ To prove
(2.13) and (2.15), we will use the identity
$${\cal R}_h^\pm(z)\left(1+h^2V{\cal R}_{0,h}^\pm(z)\right)=
{\cal R}_{0,h}^\pm(z),\quad\pm{\rm Im}\,z>0.\eqno{(2.17)}$$ Observe
that $1+h^2V{\cal R}_{0,h}^\pm(0)=1-V\Delta^{-1}$, which is supposed
to be invertible on $L^1$ with a bounded inverse denoted by $T$.
Thus, it follows from (2.12) that there exists a constant $h_0>0$ so
that for $h\ge h_0$ the operator $1+h^2V{\cal R}_{0,h}^\pm(z)$ is
invertible on $L^1$ with an inverse satisfying
$$\left\|\left(1+h^2V{\cal
R}_{0,h}^\pm(z)\right)^{-1}\right\|_{L^1\to L^1}\le C,\quad z\in
{\bf C}^\pm_\varphi,\eqno{(2.18)}$$ with a constant $C>0$
independent of $z$ and $h$. Hence, we can write
$${\cal R}_h^\pm(z)=
{\cal R}_{0,h}^\pm(z)\left(1+h^2V{\cal
R}_{0,h}^\pm(z)\right)^{-1}.\eqno{(2.19)}$$ Now (2.13) follows from
(2.11), (2.18) and (2.19), while (2.15) follows from (2.14), (2.18)
and (2.19). \eproof

Clearly, (2.1) and (2.2) follow from (2.5) and (2.14), (2.15),
respectively. To prove (2.3) we rewrite the identity (2.19) in the
form
$${\cal R}_h^\pm(z)-{\cal R}_{0,h}^\pm(z)T$$ $$=
{\cal R}_{0,h}^\pm(z)T\left(h^2V{\cal R}_{0,h}^\pm(z)-h^2V{\cal
R}_{0,h}^\pm(0)\right)T\left(1+\left(h^2V{\cal
R}_{0,h}^\pm(z)-h^2V{\cal
R}_{0,h}^\pm(0)\right)T\right)^{-1}.\eqno{(2.20)}$$ By Lemma 2.2,
(2.18) and (2.20) we conclude
$$\left\|{\cal
R}^\pm_h(z)-{\cal R}_{0,h}^\pm(z)T\right\|_{L^1\to L^1}\le
Ch^{-\beta}|{\rm Im}\,z|^{-q},\quad h\ge h_0,\,z\in {\bf
C}^\pm_\varphi, \,\,{\rm Im}\,z\neq 0,\eqno{(2.21)}$$ with constants
$C,q,\beta>0$ independent of $z$ and $h$. Now (2.3) follows from
(2.5) and (2.21). \eproof

Let $\psi_1\in C_0^\infty((0,+\infty))$, $\psi_1=1$ on supp$\,\psi$.

\begin{prop} Under the assumptions of Theorem 1.1, there exist
positive constants $h_0$ and $\beta$ so that we have the estimates
$$\int_{-\infty}^\infty\left\|Ve^{itG_0}\psi(h^2G_0)
\right\|_{L^1\to L^1}dt\le
Ch^{-\beta},\quad h\ge 1,\eqno{(2.22)}$$
$$\int_{-\infty}^\infty\left\|V\psi(h^2G)e^{itG_0}\psi_1(h^2G_0)
\right\|_{L^1\to L^1}dt\le
Ch^{-\beta},\quad h\ge h_0.\eqno{(2.23)}$$
\end{prop}

{\it Proof.} It is shown in \cite{kn:V2} (Section 2) that the kernel
of the operator $e^{itG_0}\psi(h^2G_0)$ is of the form
$K_h(|x-y|,t)$ with a function $K_h$ satisfying
$$K_h(\sigma,t)=h^{-n}K_1(\sigma h^{-1}, th^{-2}),$$
$$\left|K_1(\sigma,t)\right|\le
C|t|^{-s-1/2}\sigma^{s-(n-1)/2},\quad 0\le
s\le(n-1)/2,\,\sigma>0,\,t\neq 0.$$  Hence, for all $0\le
s\le(n-1)/2$, $\sigma>0$, $t\neq 0$, $h>0$, we have
$$\left|K_h(\sigma,t)\right|\le
Ch^{s-(n-1)/2}|t|^{-s-1/2}\sigma^{s-(n-1)/2},$$ which together with
(1.4) imply
$$\left\|Ve^{itG_0}\psi(h^2G_0)\right\|_{L^1\to L^1}\le 
Ch^{s-(n-1)/2}|t|^{-s-1/2},
\quad 1/2-\epsilon\le s\le 1/2+\epsilon,\eqno{(2.24)}$$ where
$0<\epsilon\ll 1$. Clearly, (2.22) follows from (2.24). Furthermore,
using (2.5), (2.13), (2.14) and (2.24), we get
$$\left\|V\left(\psi(h^2G)-\psi(h^2G_0)\right)e^{itG_0}\psi_1(h^2G_0)
\right\|_{L^1\to L^1}$$
 $$\le Ch^2\sum_\pm
\int_{{\bf
C^\pm_\varphi}}\left|\frac{\partial\widetilde\varphi}{\partial\bar
z}(z)\right|\left\|V{\cal R}_h^\pm(z)Ve^{itG_0}\psi_1(h^2G_0){\cal
R}_{0,h}^\pm(z)
 \right\|_{L^1\to L^1} L(dz)$$
 $$\le Ch^2\sum_\pm
\int_{{\bf
C^\pm_\varphi}}\left|\frac{\partial\widetilde\varphi}{\partial\bar
z}(z)\right|\left\|V{\cal R}_h^\pm(z)\right\|_{L^1\to
L^1}\left\|Ve^{itG_0}\psi_1(h^2G_0)\right\|_{L^1\to L^1}\left\|{\cal
R}_{0,h}^\pm(z)
 \right\|_{L^1\to L^1} L(dz)$$
 $$\le Ch^{s-(n-1)/2}|t|^{-s-1/2}\sum_\pm\int_{{\bf
C^\pm_\varphi}}\left|\frac{\partial\widetilde\varphi}{\partial\bar
z}(z)\right||{\rm Im}\,z|^{-q} L(dz)$$
 $$\le
Ch^{s-(n-1)/2}|t|^{-s-1/2},\quad 1/2-\epsilon\le s\le
1/2+\epsilon,\eqno{(2.25)}$$ which clearly implies (2.23). \eproof

\section{Proof of Theorem 1.1}

Denote
$$\Psi(t,h)=e^{itG}\psi(h^2G)-T^*e^{itG_0}\psi(h^2G_0)T,$$
 $T$ being given by (2.4). We will first show that (1.5) follows
 from the following

\begin{prop} Under the assumptions of Theorem 1.1, there exist
positive constants $C$, $h_0$ and $\beta$ so that for $h\ge h_0$ we
have
$$\left\|\Psi(t,h)\right\|_{L^1\to L^\infty}\le
Ch^{-\beta}|t|^{-n/2},\quad t\neq 0.\eqno{(3.1)}$$
\end{prop}

Recall that $\chi_a(\sigma)=\chi_1(\sigma/a)$, $a>0$ small. Then we
can write the function $\eta_a$ as follows
$$\eta_a(\sigma)=\int_{a^{-1}}^\infty\psi(\sigma\theta)\frac{d\theta}{\theta},
\quad\sigma>0,$$ where $\psi(\sigma)=\sigma\chi'_1(\sigma)\in
C_0^\infty((0,+\infty))$. Thus, we obtain from (3.1),
$$\left\|e^{itG}\eta_a(G)-T^*e^{itG_0}\eta_a(G_0)T \right\|_{L^1\to L^\infty}
\le
\int_{a^{-1}}^\infty\left\|\Psi(t,\sqrt{\theta})\right\|_{L^1\to
L^\infty}\frac{d\theta}{\theta}$$ $$\le
C|t|^{-n/2}\int_{a^{-1}}^\infty\theta^{-1-\beta/2}d\theta \le
C|t|^{-n/2},\eqno{(3.2)}$$ provided $a$ is taken small enough.
Clearly, (1.5) follows from (3.2).\\

{\it Proof of Proposition 3.1.} We will first prove the following

\begin{prop} Under the assumptions of Theorem 1.1, there exist
positive constants $C$, $h_0$ and $\beta$ so that for $h\ge h_0$ we
have
$$\int_{-\infty}^\infty\left\|Ve^{itG}\psi(h^2G)\right\|_{L^1\to L^1}dt\le
Ch^{-\beta}.\eqno{(3.3)}$$
\end{prop}

{\it Proof.} Using Duhamel's formula
$$e^{itG}=e^{itG_0}+i\int_0^te^{i(t-\tau)G}Ve^{i\tau G_0}d\tau,$$
we get the identity
$$e^{itG}\psi(h^2G)=\psi(h^2G)e^{itG_0}\psi_1(h^2G_0)T+e^{itG}\psi(h^2G)
\left(
\psi_1(h^2G)-\psi_1(h^2G_0)T\right)$$
$$+i\int_0^t\psi(h^2G)e^{i(t-\tau)G}Ve^{i\tau
G_0}\psi_1(h^2G_0)Td\tau.\eqno{(3.4)}$$ Using Propositions 2.1 and
2.3, (3.4) together with Young's inequality we obtain
$$\int_{-\infty}^\infty\left\|Ve^{itG}\psi(h^2G)\right\|_{L^1\to L^1}dt\le
Ch^{-\beta}+Ch^{-\beta}\int_{-\infty}^\infty\left\|Ve^{itG}\psi(h^2G)
\right\|_{L^1\to L^1}dt$$
$$+\int_{-\infty}^\infty\int_0^t \left\|V\psi(h^2G)e^{i(t-\tau)G}
\right\|_{L^1\to L^1}\left\|Ve^{i\tau
G_0}\psi_1(h^2G_0)\right\|_{L^1\to L^1}d\tau dt$$ $$\le
Ch^{-\beta}+Ch^{-\beta}\int_{-\infty}^\infty\left\|Ve^{itG}\psi(h^2G)
\right\|_{L^1\to L^1}dt,$$ which clearly implies (3.3) if we take
$h$ large enough. \eproof

Using Duhamel's formula
$$e^{itG}=e^{itG_0}+i\int_0^te^{i(t-\tau)G_0}Ve^{i\tau G}d\tau,$$
we get the identity
$$\Psi(t;h)=\sum_{j=1}^2\Psi_j(t;h),\eqno{(3.5)}$$
where
$$\Psi_1(t;h)=T^*\psi_1(h^2G_0)e^{itG_0}\left(\psi(h^2G)-\psi(h^2G_0)T\right)+
\left(\psi_1(h^2G)-T^*\psi_1(h^2G_0)\right)e^{itG}\psi(h^2G),$$
$$\Psi_2(t;h)=i\int_0^tT^*\psi_1(h^2G_0)e^{i(t-\tau)G_0}Ve^{i\tau
G}\psi(h^2G)d\tau.$$ By (2.1) and (2.3) together with the well known
estimate
$$\left\|e^{itG_0}\right\|_{L^1\to L^\infty}\le
C|t|^{-n/2},$$ we get
$$\left\|\Psi_1(t;h)f\right\|_{L^\infty}\le
Ch^{-\beta}|t|^{-n/2}\|f\|_{L^1}+Ch^{-\beta}\left\|\Psi(t;h)f
\right\|_{L^\infty},\quad
t\neq 0.\eqno{(3.6)}$$ By Propositions 2.3 and 3.2, $\forall f\in
L^1$, $t>0$, we have
$$t^{n/2}\left\|\Psi_2(t;h)f\right\|_{L^\infty}$$ $$\le C
\int_0^{t/2}(t-\tau)^{n/2} \left\|\psi_1(h^2G_0)e^{i(t-\tau)G_0}
\right\|_{L^1\to L^\infty}\left\|Ve^{i\tau
G}\psi(h^2G)f\right\|_{L^1}d\tau$$
 $$+C
\int_{t/2}^t\left\|\psi_1(h^2G_0)e^{i(t-\tau)G_0}V\right\|_{L^\infty\to
L^\infty}\tau^{n/2} \left\|e^{i\tau
G}\psi(h^2G)f\right\|_{L^\infty}d\tau$$
 $$\le C
\int_{-\infty}^{\infty}\left\|Ve^{i\tau
G}\psi(h^2G)f\right\|_{L^1}d\tau$$
 $$+C\sup_{t/2\le\tau\le t}\tau^{n/2} \left\|e^{i\tau
G}\psi(h^2G)f\right\|_{L^\infty}
\int_{-\infty}^{\infty}\left\|Ve^{i\tau
G_0}\psi_1(h^2G_0)\right\|_{L^1\to L^1}d\tau$$
 $$\le Ch^{-\beta}\|f\|_{L^1}+Ch^{-\beta}\sup_{t/2\le\tau\le t}\tau^{n/2}
  \left\|e^{i\tau
G}\psi(h^2G)f\right\|_{L^\infty}.\eqno{(3.7)}$$ Combining (3.5),
(3.6) and (3.7), we conclude, $\forall f\in L^1$, $t>0$,
$$t^{n/2}\left\|\Psi(t;h)f\right\|_{L^\infty}\le
Ch^{-\beta}\|f\|_{L^1}+Ch^{-\beta}t^{n/2}\left\|\Psi(t;h)f\right\|_{L^\infty}
 $$ $$+Ch^{-\beta}\sup_{t/2\le\tau\le
t}\tau^{n/2}
  \left\|\Psi(\tau;h)f\right\|_{L^\infty}.\eqno{(3.8)}$$
Taking $h$ big enough we can absorb the second and the third terms
in the RHS of (3.8), thus obtaining (3.1). Clearly, the case of
$t<0$ can be treated in the same way.\eproof

\appendix
\section{Appendix 1}

We will prove the following

\begin{lemma} Let $\psi\in C_0^\infty((0,+\infty))$. Then, 
for all $h>0$, $s\ge 0$, we
have the estimates
$$\left\|\psi(h^2G_0)\right\|_{L^1\to L^1}\le C,\eqno{(A.1)}$$
$$\left\|\langle x\rangle^s\psi(h^2G_0)\langle x\rangle^{-s}
\right\|_{L^1\to L^2}\le Ch^{-n/2}\langle h\rangle^s,\eqno{(A.2)}$$
where the constant $C$ is of the form
$$C=C'\sup_{0\le k\le
k_0}\sup_{\lambda\in{\bf
R}}|\partial_\lambda^k\psi(\lambda)|,\eqno{(A.3)}$$ with some
integer $k_0$ independent of $\psi$ and a constant $C'>0$ depending
on the support of $\psi$, only. Furthermore, if $V$ satisfies (1.1)
with $\delta>n/2$, we have the estimates (for $0<h\le 1$)
$$\left\|\psi(h^2G)-\psi(h^2G_0)\right\|_{L^1\to L^1}\le Ch^2,\eqno{(A.4)}$$
$$\left\|\langle x\rangle^\delta\left(\psi(h^2G)-\psi(h^2G_0)\right)
\right\|_{L^1\to L^2}\le Ch^{2-n/2}.\eqno{(A.5)}$$
\end{lemma}

{\it Proof.} The estimate (A.1) is proved in Section 2 using the
formula (2.5) and (2.14). It can be also seen by using the fact that
the kernel of the operator $\psi(h^2G_0)$ is of the form
$k_h(|x-y|)$ with a function $k_h$ satisfying
$$k_h(\sigma)=h^{-n}k_1(\sigma/h),\eqno{(A.6)}$$
$$|k_1(\sigma)|\le C_m\langle
\sigma\rangle^{-m},\quad\forall\sigma>0,\eqno{(A.7)}$$ for all
integers $m\ge 0$, with a constant $C_m$ of the form
$$C_m=C'_m\sup_{0\le j\le
j_m}\sup_{\lambda\in{\bf
R}}|\partial_\lambda^j\psi(\lambda)|,\eqno{(A.8)}$$ where $j_m$ is
some integer independent of $\psi$, while $C'_m>0$ depends on the
support of $\psi$. By Young's inequality, the norm in the LHS of
(A.1) is upper bounded by
$$\int_{{\bf R}^n}|k_h(|\xi|)|d\xi=\int_{{\bf
R}^n}|k_1(|\xi|)|d\xi\le C_{n+1}.$$ The norm in the LHS of (A.2) is
upper bounded by
$$\sup_{y\in{\bf R}^n}\left(\int_{{\bf R}^n}\langle x\rangle^{2s}
\langle y\rangle^{-2s}
|k_h(|x-y|)|^2dx\right)^{1/2}\le\left(\int_{{\bf R}^n}\langle
x-y\rangle^{2s} |k_h(|x-y|)|^2dx\right)^{1/2}$$
$$\le C\langle h\rangle^{s}\left(\int_{{\bf R}^n}\langle
\xi/h\rangle^{2s} |k_h(|\xi|)|^2d\xi\right)^{1/2}$$ $$=C\langle
h\rangle^{s}h^{-n/2} \left(\int_{{\bf R}^n}\langle
\xi\rangle^{2s}|k_1(|\xi|)|^2d\xi\right)^{1/2}
 \le C_{s_n}\langle h\rangle^{s}h^{-n/2},$$
 where $s_n$ is some integer depending on $n$ and $s$.
To prove (A.4) observe that by (2.5) we have
$$\psi(h^2G)-\psi(h^2G_0)=\frac{2h^2}{\pi}\int_{{\bf
C}}\frac{\partial\widetilde\varphi}{\partial\bar
z}(z)(h^2G_0-z^2)^{-1}V(h^2G-z^2)^{-1}zL(dz).\eqno{(A.9)}$$ Clearly,
(A.4) would follow from (A.9), (2.14) and the estimate (for
$z\in{\rm supp}\,\widetilde\varphi$)
$$\left\|(h^2G-z^2)^{-1}\right\|_{L^1\to L^1}\le C|{\rm Im}\,z|^{-q},
\quad 0<h\le 1,\,{\rm Im}\,z\neq 0.\eqno{(A.10)}$$ Let $\phi\in
C_0^\infty([1,2])$ be such that
$\int\phi(\theta^2)\theta^{-1}d\theta=1$. Given a parameter
$0<\varepsilon\ll 1$, we decompose the free resolvent as follows
$$(h^2G_0-z^2)^{-1}={\cal A}_\varepsilon(z;h)+{\cal
B}_\varepsilon(z;h),\eqno{(A.11)}$$ where
$${\cal A}_\varepsilon(z;h)=\int_0^1f\left((\varepsilon\theta
h)^2G_0;(\varepsilon\theta)^2;z\right)\frac{d\theta}{\theta},$$
$${\cal B}_\varepsilon(z;h)=\int_1^\infty f\left((\varepsilon\theta
h)^2G_0;(\varepsilon\theta)^2;z\right)\frac{d\theta}{\theta},$$
where
$$f(\lambda;\mu;z)=\frac{\phi(\lambda)}{\lambda\mu^{-1}-z^2}.$$
It is easy to see that there exist constants $0<\mu_1<\mu_2$ so that
the function $f$ satisfies the following bounds
$$\left|\partial_\lambda^jf(\lambda;\mu;z)\right|\le C_j\mu, \quad
0<\mu\le\mu_1,\eqno{(A.12)}$$
$$\left|\partial_\lambda^jf(\lambda;\mu;z)\right|\le C'_j
|{\rm Im}\,z|^{-j-1}, \quad
\mu_1\le\mu\le\mu_2,\eqno{(A.13)}$$
$$\left|\partial_\lambda^jf(\lambda;\mu;z)\right|\le C''_j, \quad
\mu\ge\mu_2,\eqno{(A.14)}$$ for every integer $j\ge 0$. By (A.1),
(A.3) and (A.12), we have
$$\left\|f\left((\varepsilon\theta
h)^2G_0;(\varepsilon\theta)^2;z\right)\right\|_{L^1\to L^1}\le
C(\varepsilon\theta)^2,\quad 0<\theta\le 1 ,\eqno{(A.15)}$$ provided
$\varepsilon>0$ is taken small enough. We deduce from (A.15),
$$\left\|{\cal A}_\varepsilon(z;h)\right\|_{L^1\to L^1}\le
C\varepsilon^2,\quad z\in{\rm
supp}\,\widetilde\varphi,\eqno{(A.16)}$$ with a constant $C>0$
independent of $z$, $h$ and $\varepsilon$. By (A.2), (A.3),
(A.12)-(A.14), we have
$$\left\|\langle x\rangle^s f\left((\varepsilon\theta
h)^2G_0;(\varepsilon\theta)^2;z\right)\langle
x\rangle^{-s}\right\|_{L^1\to L^2}\le C(\varepsilon\theta
h)^{-n/2}\langle \varepsilon\theta h\rangle^s |{\rm
Im}\,z|^{-q},\eqno{(A.17)}$$ with constants $C,q>0$ independent of
$z$, $\theta$, $h$ and $\varepsilon$. We deduce from (A.17),
$$\left\|{\cal B}_\varepsilon(z;h)\right\|_{L^1\to L^2}\le
C'_\varepsilon h^{-n/2}|{\rm
Im}\,z|^{-q}\int_1^\infty\theta^{-1-n/2}d\theta$$
 $$\le C_\varepsilon h^{-n/2}|{\rm
Im}\,z|^{-q},\quad z\in{\rm supp}\,\widetilde\varphi,\eqno{(A.18)}$$
with a constant $C_\varepsilon >0$ independent of $z$ and $h$. It
follows from (A.16) that the operator $1+h^2V{\cal
A}_\varepsilon(z;h)$ is invertible on $L^1$, provided
$\varepsilon>0$ is taken small enough, independent of $h$.
Therefore, we can write the identity
$$\left(h^2G-z^2\right)^{-1}=\left(h^2G_0-z^2\right)^{-1}
+h^2\left(h^2G-z^2\right)^{-1}
V\left(h^2G_0-z^2\right)^{-1},\eqno{(A.19)}$$ in the form
$$\left(h^2G-z^2\right)^{-1}={\cal M}(z;h)+h^2\left(h^2G-z^2\right)^{-1}
{\cal N}(z;h),\eqno{(A.20)}$$ where the operators
$${\cal M}(z;h)=\left(h^2G_0-z^2\right)^{-1}\left(1+h^2V{\cal
A}_\varepsilon(z;h)\right)^{-1},$$
$${\cal N}(z;h)=V{\cal B}_\varepsilon(z;h)\left(1+h^2V{\cal
A}_\varepsilon(z;h)\right)^{-1},$$ satisfy the estimates
$$\left\|{\cal M}(z;h)\right\|_{L^1\to L^1}+\left\|{\cal N}(z;h)
\right\|_{L^1\to L^1}
\le C|{\rm Im}\,z|^{-q},\eqno{(A.21)}$$
$$\left\|{\cal N}(z;h)\right\|_{L^1\to L^2}
\le Ch^{-n/2}|{\rm Im}\,z|^{-q}.\eqno{(A.22)}$$ By (A.20) we have
$$\left(h^2G-z^2\right)^{-1}=\sum_{j=0}^{J-1}{\cal M}(z;h){\cal N}(z;h)^j
+h^{2J}\left(h^2G-z^2\right)^{-1} {\cal N}(z;h)^J$$
$$=\sum_{j=0}^{J-1}{\cal M}(z;h){\cal N}(z;h)^j
+h^{2J}\left(h^2G_0-z^2\right)^{-1} {\cal N}(z;h)^J$$ $$
+h^{2J+2}\left(h^2G_0-z^2\right)^{-1}V
\left(h^2G-z^2\right)^{-1}{\cal N}(z;h)^J,\eqno{(A.23)}$$ for every
integer $J\ge 1$.  By (A.22) and (2.14), we obtain
$$\left\|\left(h^2G_0-z^2\right)^{-1}V
\left(h^2G-z^2\right)^{-1}{\cal N}(z;h)\right\|_{L^1\to L^1}$$
 $$\le \|V\|_{L^2}\left\|\left(h^2G_0-z^2\right)^{-1}\right\|_{L^1\to L^1}
 \left\|\left(h^2G-z^2\right)^{-1}\right\|_{L^2\to L^2}
 \left\|{\cal N}(z;h)\right\|_{L^1\to L^2}$$ $$
\le Ch^{-n/2}|{\rm Im}\,z|^{-q_2}.\eqno{(A.24)}$$ Now, (A.10)
follows from (A.21), (A.23) and (A.24).

To prove (A.5) we rewrite (A.20) in the form
$$\left(h^2G-z^2\right)^{-1}-\left(h^2G_0-z^2\right)^{-1}=\sum_{j=1}^3
{\cal F}_j(z;h),\eqno{(A.25)}$$ where
$${\cal F}_1(z;h)=h^2{\cal
A}_\varepsilon(z;h)V{\cal A}_\varepsilon(z;h)\left(1+h^2V{\cal
A}_\varepsilon(z;h)\right)^{-1},$$
$${\cal F}_2(z;h)=h^2{\cal
B}_\varepsilon(z;h)V{\cal A}_\varepsilon(z;h)\left(1+h^2V{\cal
A}_\varepsilon(z;h)\right)^{-1},$$
$${\cal F}_3(z;h)=h^2\left(h^2G-z^2\right)^{-1}
V{\cal B}_\varepsilon(z;h)\left(1+h^2V{\cal
A}_\varepsilon(z;h)\right)^{-1}.$$ It is easy to see that we have
the estimate
$$\left\|\langle x\rangle^s\left(h^2G-z^2\right)^{-1}
\langle x\rangle^{-s}\right\|_{L^2\to L^2}\le  C|{\rm
Im}\,z|^{-q},\quad z\in{\rm supp}\,\widetilde\varphi,\,0< h\le
1,\eqno{(A.26)}$$ for every $s\ge 0$ with constants $C,q>0$
depending on $s$ but independent of $z$ and $h$. By (A.18) and
(A.26),
$$\left\|\langle x\rangle^\delta{\cal F}_3(z;h)\right\|_{L^1\to L^2}\le
Ch^{2-n/2}|{\rm Im}\,z|^{-q},\quad z\in{\rm
supp}\,\widetilde\varphi,\,0< h\le 1.\eqno{(A.27)}$$ Observe now
that we can write the operator ${\cal A}_\varepsilon(z;h)$ in the
form
$${\cal
A}_\varepsilon(z;h)=\chi^{(3)}_\varepsilon(h^2G_0)\left(h^2G_0-z^2
\right)^{-1},$$
where
$$\chi^{(3)}_\varepsilon(\sigma)=\int_0^{\varepsilon\sigma^{1/2}}
\phi(\theta^2)\frac{d\theta}
{\theta}.$$ Similarly, we can decompose the operator ${\cal
B}_\varepsilon(z;h)$ as  ${\cal B}^{(1)}_\varepsilon+{\cal
B}^{(2)}_\varepsilon$, where
$${\cal B}^{(j)}_\varepsilon(z;h)=\chi^{(j)}_\varepsilon(h^2G_0)
\left(h^2G_0-z^2\right)^{-1},\quad j=1,2,$$
$$\chi^{(1)}_\varepsilon(\sigma)=\int_{\varepsilon\sigma^{1/2}}^{
\varepsilon^{-1}\sigma^{1/2}}\phi(\theta^2)\frac{d\theta}
{\theta},\quad \chi^{(2)}_\varepsilon(\sigma)=\int_{
\varepsilon^{-1}\sigma^{1/2}}^\infty\phi(\theta^2)\frac{d\theta}
{\theta}.$$ Taking $\varepsilon>0$ small enough we can arrange that
 supp$\,\chi^{(j)}_\varepsilon\cap\,$supp$\,\varphi=\emptyset$, $j=2,3$, so
the operator-valued functions ${\cal A}_\varepsilon(z;h)$ and ${\cal
B}^{(2)}_\varepsilon(z;h)$ are analytic on
supp$\,\widetilde\varphi$. Therefore, we can write (A.9) in the form
$$\psi(h^2G)-\psi(h^2G_0)=\frac{2}{\pi}\sum_{j=3}^4\int_{{\bf
C}}\frac{\partial\widetilde\varphi}{\partial\bar z}(z){\cal
F}_j(z;h)zL(dz),\eqno{(A.28)}$$ where
$${\cal F}_4(z;h)=h^2{\cal
B}^{(1)}_\varepsilon(z;h)V{\cal A}_\varepsilon(z;h)\left(1+h^2V{\cal
A}_\varepsilon(z;h)\right)^{-1}.$$ By (A.17) (with $s=\delta$), we
have
$$\left\|\langle x\rangle^\delta{\cal F}_4(z;h)\right\|_{L^1\to L^2}
\le Ch^{2}\left\|\langle x\rangle^\delta {\cal
B}^{(1)}_\varepsilon(z;h)\langle x\rangle^{-\delta}\right\|_{L^1\to
L^2}$$  $$
 \le Ch^{2-n/2}|{\rm Im}\,z|^{-q},\quad z\in{\rm
supp}\,\widetilde\varphi,\,0< h\le 1.\eqno{(A.29)}$$ Now (A.5)
follows from (A.27)-(A.29). \eproof

\section{Appendix 2}

 Combining some ideas from \cite{kn:V1},\cite{kn:V2} and \cite{kn:JSS}
 we will prove the following

\begin{Theorem}  Let $n\ge 4$, let $V$ satisfy (1.1) with $\delta>n-1$
as well as (1.3). Then, for every $a>0$ we have the estimate
$$\left\|e^{itG}\chi_a(G)\right\|_{L^1\to L^\infty}\le
C|t|^{-n/2},\quad t\neq 0.\eqno{(B.1)}$$
\end{Theorem}

\noindent
 {\bf Remark.} Note that (B.1) is proved in \cite{kn:JSS} for
 potentials satisfying (1.1) with $\delta>n$, the condition (1.3) as
 well as an extra technical assumption. Here we eliminate this extra
 assumption.

 {\it Proof.} The key point in the proof in \cite{kn:JSS} is the
 bound
 $$\left\|e^{-itG_0}Ve^{itG_0}\right\|_{L^1\to L^1}\le\|\widehat
 V\|_{L^1},\quad\forall t.\eqno{(B.2)}$$ Combining (B.2) with
 Duhamel's formula one easily gets
 $$\left\|e^{-itG_0}Ve^{itG}\right\|_{L^1\to L^1}\le
 C,\quad |t|\le 1,\eqno{(B.3)}$$ with a constant $C>0$ independent
 of $t$. In what follows we will derive (B.1) from (B.2) and (B.3).
 To this end, given a function $\psi\in C_0^\infty((0,+\infty))$ and
 a parameter $0<h\le 1$, as in \cite{kn:V1}, \cite{kn:V2}, denote
$$\Psi(t,h)=e^{itG}\psi(h^2G)-e^{itG_0}\psi(h^2G_0),$$ 
 $$F(t)=i\int_0^te^{i(t-\tau)G_0}Ve^{i\tau G_0}d\tau.$$ As in these papers, 
it is easy to see that (B.1) follows from the following

\begin{Theorem}  Under the assumptions of Theorem B.1, there exist constants 
$C,\beta>0$
so that we have the estimates (for $0<h\le 1$, $t\neq 0$)
$$\left\|F(t)\right\|_{L^1\to L^\infty}\le
C|t|^{-n/2},\eqno{(B.4)}$$
$$\left\|\Psi(t,h)-F(t)\psi(h^2G_0)\right\|_{L^1\to L^\infty}\le
Ch^\beta|t|^{-n/2}.\eqno{(B.5)}$$
\end{Theorem}

{\it Proof.} Clearly, (B.4) follows from (B.2) for $|t|\le 2$. Let $|t|\ge 2$.
Without loss of generality we may suppose $t\ge 2$. 
Write $F=F_1+F_2$, where
 $$F_1(t)=i\int_1^{t-1}e^{i(t-\tau)G_0}Ve^{i\tau G_0}d\tau,$$
 $$F_2(t)=i\left(\int_0^1+\int_{t-1}^t\right)
e^{i(t-\tau)G_0}Ve^{i\tau G_0}d\tau.$$
It follows from (B.2) that $F_2(t)$ satisfies (B.4). To deal with the operator
$F_1(t)$, observe that its kernel is of the form
$$c_n\int_{{\bf R}^n} U(|x-\xi|^2/4,|y-\xi|^2/4,t)V(\xi)d\xi,$$
where $c_n$ is a constant and
$$U(\sigma_1,\sigma_2,t)=\int_1^{t-1}e^{i\sigma_1/(t-\tau)
+i\sigma_2/\tau}(t-\tau)^{-n/2}\tau^{-n/2}d\tau.$$
To prove that $F_1(t)$ satisfies (B.4), it suffices to show that
$$\left|U(\sigma_1,\sigma_2,t)\right|\le Ct^{-n/2}\left(\sigma_1^{-1/2}+
\sigma_2^{-1/2}\right),\quad\forall\sigma_1,\sigma_2>0,\,t\ge 2.
\eqno{(B.6)}$$
To do so, observe that
$$U(\sigma_1,\sigma_2,t)=t^{-n+1}\left(u(\sigma_1t^{-1},\sigma_2t^{-1},
t^{-1})+u(\sigma_2t^{-1},\sigma_1t^{-1},t^{-1})\right),\eqno{(B.7)}$$
where
$$u(\sigma'_1,\sigma'_2,\kappa)=\int_\kappa^{1/2}e^{i\sigma'_1/(1-\tau')
+i\sigma'_2/\tau'}(1-\tau')^{-n/2}(\tau')^{-n/2}d\tau'.$$
It is easy to see that (B.6) follows from (B.7) and the bound
$$\left|u(\sigma'_1,\sigma'_2,\kappa)\right|\le C\kappa^{-(n-3)/2}
(\sigma'_2)^{-1/2},\quad\forall \sigma'_1,\sigma'_2>0,\,0<\kappa\le 1/2.
\eqno{(B.8)}$$
To prove (B.8), we make a change of variables $\mu=1/\tau'$ and write the
function $u$ in the form
$$u(\sigma'_1,\sigma'_2,\kappa)=\int_2^{\kappa^{-1}}e^{i\varphi(\mu,
\sigma'_1,\sigma'_2)}\left(\frac{\mu}{\mu-1}\right)^{n/2}\mu^{n/2-2}d\mu,$$
where
$$\varphi(\mu,\sigma'_1,\sigma'_2)=\mu\sigma'_2+\frac{\mu}{\mu-1}\sigma'_1.$$
We have
$$\left|u(\sigma'_1,\sigma'_2,\kappa)\right|\le 
C\int_2^{\kappa^{-1}}\mu^{n/2-2}d\mu\le C\kappa^{-(n-2)/2}.\eqno{(B.9)}$$
Furthermore, observe that
$$\varphi'(\mu)=\frac{d\varphi}{d\mu}=\sigma'_2-\frac{\sigma'_1}{(\mu-1)^2},$$
so $\varphi'$ vanishes at $\mu_0=1+(\sigma'_1/\sigma'_2)^{1/2}$. We will consider now two cases.

Case 1. $\mu_0\not\in [3/2,3\kappa^{-1}/2]$. Then, we have
$$\left|\varphi'(\mu)\right|\ge \sigma'_2\frac{|\mu-\mu_0|}{\mu-1}\ge \frac{\sigma'_2}{10},\quad \mu \in [2,\kappa^{-1}].$$
Therefore, integrating by parts, we obtain
$$u(\sigma'_1,\sigma'_2,\kappa)=\int_2^{\kappa^{-1}}(i\varphi')^{-1}
\left(\frac{\mu}{\mu-1}\right)^{n/2}\mu^{n/2-2}de^{i\varphi}$$
$$=\left.e^{i\varphi}(i\varphi')^{-1}
\left(\frac{\mu}{\mu-1}\right)^{n/2}\mu^{n/2-2}\right|_2^{\kappa^{-1}}-
\int_2^{\kappa^{-1}}e^{i\varphi}f(\mu)d\mu,\eqno{(B.10)}$$
where
$$f(\mu)=\frac{d}{d\mu}\left((i\varphi')^{-1}
\left(\frac{\mu}{\mu-1}\right)^{n/2}\mu^{n/2-2}\right)$$ $$=
(i\varphi')^{-1}\frac{d}{d\mu}\left(
\left(\frac{\mu}{\mu-1}\right)^{n/2}\mu^{n/2-2}\right)+
\frac{i\varphi''}{\varphi'^2}
\left(\frac{\mu}{\mu-1}\right)^{n/2}\mu^{n/2-2}.$$
Since
$$\left|\frac{\varphi''}{\varphi'}\right|\le\frac{2\sigma'_1
(\mu-1)^{-2}}{(\mu-1)\left|\sigma'_2-\sigma'_1(\mu-1)^{-2}\right|}
\le \frac{2}{\mu-1}\left(1+\frac{\sigma'_2}{|\varphi'|}\right)\le \frac{22}{\mu-1},$$
we have (for $\mu\ge 2$)
$$|f(\mu)|\le C(\sigma'_2)^{-1}\mu^{n/2-3}.\eqno{(B.11)}$$
By (B.10) and (B.11),
$$\left|u(\sigma'_1,\sigma'_2,\kappa)\right|\le 
 C(\sigma'_2)^{-1}\kappa^{-(n-4)/2}.\eqno{(B.12)}$$
Clearly, in this case (B.8) follows from (B.9) and (B.12).

Case 2. $\mu_0\in [3/2,3\kappa^{-1}/2]$. Denote $I(\mu_0)=[9\mu_0/10,11\mu_0/10]\cap [2,\kappa^{-1}]$. We write the function $u$
as $u_1+u_2$, where
$$u_1=\int_{I(\mu_0)}e^{i\varphi}\left(\frac{\mu}{\mu-1}\right)^{n/2}\mu^{n/2-2}d\mu
=\mu_0^{n/2-1}\int_{\widetilde I(\mu_0)}e^{i\lambda\phi(z)}g(z)dz,\eqno{(B.13)}$$
where we have made a change of variables $\mu=\mu_0(1+z)$, $\widetilde I(\mu_0)\subset [-1/10,1/10]$, $\lambda=\mu_0\sigma'_2$,
$$g(z)=\left(\frac{1+z}{1+z-\mu_0^{-1}}\right)^{n/2}(1+z)^{n/2-2},$$
$$\phi(z)=(1+z)\left(1+\frac{(\mu_0-1)^2}{\mu_0(1+z)-1}\right)=\mu_0+\frac{\mu_0}{\mu_0-1}z^2+O(z^3),\quad |z|\ll 1,$$
uniformly in $\mu_0$. It is easy to see that we have the estimate
$$\left|\int_0^a e^{i\lambda\phi(z)}g(z)dz\right|\le C\lambda^{-1/2},\quad |a|\le 1/10.\eqno{(B.14)}$$
Indeed, the functions $g(z)$ and $\phi(z)$ are analytic in $|z|\le 1/10$ with $|g(z)|$ bounded there uniformly
in $\mu_0$. Therefore, we can change the contour of integration to obtain (with some $0<\gamma\ll 1$)
$$\left|\int_0^a e^{i\lambda\phi(z)}g(z)dz\right|\le \left|\int_0^a e^{i\lambda\phi(e^{i\gamma}y)}g(e^{i\gamma}y)dy\right|
+\left|a\int_0^\gamma e^{i\lambda\phi(e^{i\theta}a)}g(e^{i\theta}a)d\theta\right|$$ $$\le
C_1\int_0^a e^{-C\lambda y^2}dy+C'_1\int_0^\gamma e^{-C'\lambda\theta}d\theta=O(\lambda^{-1/2}),$$
with some constants $C,C',C_1,C'_1>0$. By (B.13) and (B.14) we conclude
$$|u_1|\le C(\sigma'_2)^{-1/2}\mu_0^{(n-3)/2}\le\widetilde C(\sigma'_2)^{-1/2}\kappa^{-(n-3)/2}.\eqno{(B.15)}$$
On the other hand, if $\mu\in[2,\kappa^{-1}]\setminus I(\mu_0)$, then $$\frac{|\mu-\mu_0|}{\mu-1}\ge C>0,$$
so we can bound from below $|\varphi'(\mu)|$. Therefore, the function $u_2$ can be treated in the same way as does $u$ in Case 1.
Thus, $u_2$ satisfies (B.8) and hence, in view of (B.15), so does $u$. This completes the proof of (B.4).

It suffices to prove (B.5) for $0<h\le h_0$ with some
constant $0<h_0\le 1$, since for $h_0\le h\le 1$ it follows from (B.4) and
the estimate of the $L^1\to L^\infty$ norm of $\Psi(t,h)$ proved in
\cite{kn:V2} for the larger class of potentials satisfying (1.1)
with $\delta>(n+2)/2$ (without using (1.3)). Without loss
of generality we may suppose $t>0$. Now, using Duhamel's formula as
in \cite{kn:V1}, \cite{kn:V2} we get the identity
$$\Psi(t;h)-F(t)\psi(h^2G_0)=\sum_{j=1}^5\Psi_j(t;h),\eqno{(B.16)}$$
where
$$\Psi_1(t;h)=\psi_1(h^2G_0)e^{itG_0}\left(\psi(h^2G)-\psi(h^2G_0)\right)$$ 
$$+
\left(\psi_1(h^2G)-\psi_1(h^2G_0)\right)e^{itG_0}\psi(h^2G_0)+
\left(\psi_1(h^2G)-\psi_1(h^2G_0)\right)\Psi(t;h),$$
$$\Psi_2(t;h)=i\left(\int_0^\gamma+\int_{t-\gamma}^t\right)
\psi_1(h^2G_0)e^{i(t-\tau)G_0}Ve^{i\tau
G}\psi(h^2G)d\tau,$$
$$\Psi_3(t;h)=-i\left(\int_0^\gamma+\int_{t-\gamma}^t\right)
e^{i(t-\tau)G_0}Ve^{i\tau
G_0}\psi(h^2G_0)d\tau,$$
$$\Psi_4(t;h)=i\int_\gamma^{t-\gamma}\psi_1(h^2G_0)e^{i(t-\tau)G_0}V
\Psi(\tau;h)d\tau,$$ 
$$\Psi_5(t;h)=-i\int_\gamma^{t-\gamma}(1-\psi_1)(h^2G_0)e^{i(t-\tau)G_0}V
e^{i\tau G_0}\psi(h^2G_0)d\tau,$$
where $\psi_1\in C_0^\infty((0,+\infty))$,
$\psi_1=1$ on supp$\,\psi$, and $0<\gamma\ll 1$ is a parameter to be
fixed later on, depending on $h$. In view of (A.4), we have
$$\left\|\Psi_1(t;h)f\right\|_{L^\infty}\le
Ch^2t^{-n/2}\|f\|_{L^1}+Ch^2\left\|\Psi(t;h)f\right\|_{L^\infty},\quad
\forall f\in L^1.\eqno{(B.17)}$$ By (B.2) and (B.3),
$$\left\|\Psi_j(t;h)f\right\|_{L^\infty}\le
C\gamma
t^{-n/2}\|f\|_{L^1}+C\gamma\left\|\Psi(t;h)f\right\|_{L^\infty},\quad
\forall f\in L^1,\,j=2,3,\eqno{(B.18)}$$ with a constant $C>0$ independent of
$t$, $h$ and $\gamma$. 

\begin{prop} Let $V$ satisfy (1.1) with $\delta>n-1$. Then,
there exist constants $C,\beta_1>0$ 
so that for $0<h\le 1$, $t\ge 2\gamma$, we have the estimate
$$\left\|\Psi_4(t,h)\right\|_{L^1\to L^\infty}\le
Ch^{\beta_1}\gamma^{-(n-3)/2}t^{-n/2}.\eqno{(B.19)}$$
\end{prop}

 {\it Proof.} We will make use of
 the following estimates proved in \cite{kn:V2}.

\begin{prop} Let $V$ satisfy (1.1) with $\delta>(n+2)/2$. Then,
for every $0<\epsilon\ll 1$, $1/2-\epsilon/4\le s\le (n-1)/2$, 
$0<h\le 1$, $t\neq 0$, we have the estimates
$$\left\|\psi(h^2G_0)e^{itG_0}\langle x\rangle^{-s-1/2-\epsilon}
\right\|_{L^2\to L^\infty}\le
Ch^{s-(n-1)/2}|t|^{-s-1/2},\eqno{(B.20)}$$
$$\left\|\Psi(t,h)\langle x\rangle^{-s-1/2-\epsilon}
\right\|_{L^2\to L^\infty}\le
Ch^{s-(n-3)/2-\epsilon/4}|t|^{-s-1/2}.\eqno{(B.21)}$$
\end{prop}

By (B.20) and (B.21), we get (with some
$0<\varepsilon_0\ll 1$)
$$\left\|\Psi_4(t,h)\right\|_{L^1\to L^\infty}$$ $$\le
C\int_\gamma^{t/2}\left\|\psi_1(h^2G_0)e^{i(t-\tau)G_0}\langle
x\rangle^{-n/2-\varepsilon_0} \right\|_{L^2\to L^\infty}\left\|
\langle
x\rangle^{-(n-2)/2-\varepsilon_0}\Psi(\tau,h)\right\|_{L^1\to
L^2}d\tau$$
$$+
C\int^{t-\gamma}_{t/2}\left\|\psi_1(h^2G_0)e^{i(t-\tau)G_0}\langle
x\rangle^{-(n-2)/2-\varepsilon_0} \right\|_{L^2\to L^\infty}\left\|
\langle x\rangle^{-n/2-\varepsilon_0}\Psi(\tau,h)\right\|_{L^1\to
L^2}d\tau$$
$$\le Ch^{\varepsilon_0/4}t^{-n/2}\int_\gamma^{t/2}\tau^{-(n-2)/2}d\tau
+Ch^{\varepsilon_0/4}t^{-n/2}\int^{t-\gamma}_{t/2}(t-\tau)^{-(n-2)/2}d\tau
$$ $$\le Ch^{\varepsilon_0/4}\gamma^{-(n-3)/2}t^{-n/2}.$$
\eproof

\begin{prop} Let $V$ satisfy (1.1) with $\delta>n-1$. Then,
for every $0<\epsilon\ll 1$,  $0<h\le 1$, $t\ge 2\gamma$, we have the estimate
$$\left\|\Psi_5(t,h)\right\|_{L^1\to L^\infty}\le
C_\epsilon h^{\epsilon}\gamma^{-(n-3)/2-\epsilon}t^{-n/2}.\eqno{(B.22)}$$
\end{prop}

{\it Proof.} We will make use of the fact that the kernel of the
operator $e^{itG_0}\psi(h^2G_0)$ is of the form
$K_h(|x-y|,t)$, where
$$K_h(\sigma,t)=\frac{\sigma^{-2\nu}}{(2\pi)^{\nu +1}}\int_0^\infty
e^{it\lambda^2}{\cal J}_\nu(\sigma\lambda)\psi(h^2\lambda^2)\lambda
d\lambda=h^{-n}K_1(\sigma h^{-1},th^{-2}),\eqno{(B.23)}$$ where
${\cal J}_\nu(z)=z^\nu J_\nu(z)$,
$J_\nu(z)=\left(H_\nu^+(z)+H_\nu^-(z)\right)/2$ is the Bessel
function of order $\nu=(n-2)/2$. So, the kernel of the operator
$\Psi_5$ is of the form
$$\int_{{\bf R}^n}W_h(|x-\xi|,|y-\xi|,t,\gamma)V(\xi)d\xi,$$
where
$$W_h(\sigma_1,\sigma_2,t,\gamma)=-i\int_\gamma^{t-\gamma}\widetilde
K_h(\sigma_1,t-\tau)K_h(\sigma_2,\tau)d\tau$$ $$=h^{-2n+2}W_1(\sigma_1h^{-1},
\sigma_2
h^{-1},th^{-2},\gamma h^{-2}),\eqno{(B.24)}$$ where $\widetilde K_h$
is defined by replacing in the definition of $K_h$ the function
$\psi$ by $1-\psi_1$. It is easy to see that (B.22) follows from the
bound (for all $\sigma_1$, $\sigma_2$, $\gamma>0$, $0<\epsilon\ll
1$, $t\ge 2\gamma$)
$$\left|W_h(\sigma_1,\sigma_2,t,\gamma)\right|\le C_\epsilon h^{\epsilon}
\gamma^{-(n-3)/2-\epsilon}t^{-n/2}\left(\sigma_1^{-n+2}+
\sigma_1^{-1+\epsilon}+\sigma_2^{-n+2}
+\sigma_2^{-1+\epsilon}\right).\eqno{(B.25)}$$ In
view of (B.24), it suffices to prove (B.25) with $h=1$. Now, observe
that $W_1=W_1^{(1)}-W_1^{(2)}$, where
$$W_1^{(1)}(\sigma_1,\sigma_2,t,\gamma)=\frac{(\sigma_1\sigma_2)^{-2\nu}}
{4(2\pi)^{2\nu +2}}\int_0^\infty\int_0^\infty
e^{i(t-\gamma)\lambda_1^2+i\gamma\lambda_2^2}\rho(\lambda_1^2,\lambda_2^2)
{\cal J}_\nu(\sigma_1\lambda_1){\cal
J}_\nu(\sigma_2\lambda_2)
d\lambda_1^2d\lambda_2^2,$$
$$W_1^{(2)}(\sigma_1,\sigma_2,t,\gamma)=\frac{(\sigma_1\sigma_2)^{-2\nu}}
{4(2\pi)^{2\nu +2}}\int_0^\infty\int_0^\infty
e^{i(t-\gamma)\lambda_2^2+i\gamma\lambda_1^2}\rho(\lambda_1^2,\lambda_2^2)
{\cal J}_\nu(\sigma_1\lambda_1){\cal
J}_\nu(\sigma_2\lambda_2)
d\lambda_1^2d\lambda_2^2,$$
where the function
$$\rho(\lambda_1^2,\lambda_2^2)=\frac{(1-\psi_1)(\lambda_1^2)\psi(\lambda_2^2)}
{\lambda_2^2-\lambda_1^2}=(1-\psi_1)(\lambda_1^2)\psi_1(\lambda_2^2)\frac{
\psi(\lambda_2^2)-\psi(\lambda_1^2)}
{\lambda_2^2-\lambda_1^2}$$
satisfies the bound
$$\left|\partial_{\lambda_1^2}^{\alpha_1}\partial_{\lambda_2^2}^{\alpha_2}
\rho(\lambda_1^2,\lambda_2^2)\right|\le C_{\alpha_1,\alpha_2}\langle
 \lambda_1^2\rangle^{-1-\alpha_1},\quad\forall(\lambda_1,\lambda_2).
\eqno{(B.26)}$$
Given any integers $0\le k,m<n/2$, since ${\cal J}_\nu(z)=O(z^{n-2})$
as $z\to 0$, we can integrate by parts to get
$$W_1^{(1)}(\sigma_1,\sigma_2,t,\gamma)=i^{-m-k}(t-\gamma)^{-k}\gamma^{-m}
\frac{(\sigma_1\sigma_2)^{-2\nu}}
{4(2\pi)^{2\nu +2}}\int_0^\infty\int_0^\infty
e^{i(t-\gamma)\lambda_1^2+i\gamma\lambda_2^2}$$ $$\times
\partial_{\lambda_1^2}^k
\partial_{\lambda_2^2}^m\left(\rho(\lambda_1^2,\lambda_2^2)
{\cal J}_\nu(\sigma_1\lambda_1){\cal J}_\nu(\sigma_2\lambda_2)\right)
d\lambda_1^2d\lambda_2^2,$$
$$W_1^{(2)}(\sigma_1,\sigma_2,t,\gamma)=i^{-m-k}(t-\gamma)^{-k}\gamma^{-m}
\frac{(\sigma_1\sigma_2)^{-2\nu}}
{4(2\pi)^{2\nu +2}}\int_0^\infty\int_0^\infty
e^{i(t-\gamma)\lambda_2^2+i\gamma\lambda_1^2}$$ $$\times
\partial_{\lambda_1^2}^m
\partial_{\lambda_2^2}^k\left(\rho(\lambda_1^2,\lambda_2^2)
{\cal J}_\nu(\sigma_1\lambda_1){\cal J}_\nu(\sigma_2\lambda_2)\right)
d\lambda_1^2d\lambda_2^2.$$
Using the inequality
$$\left|\int_{-\infty}^\infty e^{it\lambda^2}\varphi(\lambda)d\lambda\right|
\le C|t|^{-1/2}\left\|\widehat\varphi\right\|_{L^1},\eqno{(B.27)}$$
we obtain (for $t\ge 2\gamma$)
$$\left|W_1^{(1)}(\sigma_1,\sigma_2,t,\gamma)\right|\le Ct^{-k-1/2}
\gamma^{-m}(\sigma_1\sigma_2)^{-2\nu}\int_{-\infty}^\infty\left|
\int_0^\infty\int_0^\infty
e^{i\tau\lambda_1+i\gamma\lambda_2^2}\right.$$ $$\left.\times \lambda_1
\partial_{\lambda_1^2}^k
\partial_{\lambda_2^2}^m\left(\rho(\lambda_1^2,\lambda_2^2)
{\cal J}_\nu(\sigma_1\lambda_1){\cal J}_\nu(\sigma_2\lambda_2)\right)
d\lambda_1d\lambda_2^2\right|d\tau,\eqno{(B.28)}$$
$$\left|W_1^{(2)}(\sigma_1,\sigma_2,t,\gamma)\right|\le Ct^{-k-1/2}
\gamma^{-m}(\sigma_1\sigma_2)^{-2\nu}\int_{-\infty}^\infty\left|
\int_0^\infty\int_0^\infty
e^{i\tau\lambda_2+i\gamma\lambda_1^2}\right.$$ $$\left.\times \lambda_2
\partial_{\lambda_1^2}^m
\partial_{\lambda_2^2}^k\left(\rho(\lambda_1^2,\lambda_2^2)
{\cal J}_\nu(\sigma_1\lambda_1){\cal J}_\nu(\sigma_2\lambda_2)\right)
d\lambda_2d\lambda_1^2\right|d\tau.\eqno{(B.29)}$$
Recall now that the function ${\cal J}_\nu$ is of the form
${\cal J}_\nu(z)=e^{iz}b_\nu^+(z)+e^{-iz}b_\nu^-(z)$, where
$b_\nu^\pm(z)$ are symbols of order $(n-3)/2$ for $z\ge 1$, while
near $z=0$ the function ${\cal J}_\nu(z)$ is equal to $z^{2\nu}$ times
an analytic function. Therefore, it satisfies the bounds
$$\left|\partial_z^j{\cal J}_\nu(z)\right|\le Cz^{n-2-j}
\langle z\rangle^{j-(n-1)/2},\quad\forall z>0,\,0\le j\le n-2,\eqno{(B.30)}$$
$$\left|\partial_z^j{\cal J}_\nu(z)\right|\le C_j
\langle z\rangle^{(n-3)/2},\quad\forall z>0,\,j\ge 0.\eqno{(B.31)}$$
Moreover, the functions $b_\nu^\pm(z)$ are of the form (near $z=0$)
$$b_\nu^\pm(z)=b_{\nu,1}^\pm(z)+z^{n-2}\log z\,b_{\nu,2}^\pm(z),$$
where $b_{\nu,j}^\pm(z)$ are analytic functions, $b_{\nu,2}^\pm(z)\equiv 0$
if $n$ is odd. Therefore, we have
$$\left|\partial_z^jb_{\nu}^\pm(z)\right|\le C,\quad 0<z\le 1,\, 0\le j\le
 n-3,$$
 $$\left|\partial_z^jb_{\nu}^\pm(z)\right|\le C_\epsilon z^{-\epsilon},
\quad 0<z\le 1,\, j=n-2,$$
  $$\left|\partial_z^jb_{\nu}^\pm(z)\right|\le C_jz^{n-2-j},
\quad 0<z\le 1,\, j\ge n-1,$$
which imply
$$\left|\partial_z^jb_{\nu}^\pm(z)\right|\le C\langle z\rangle^{(n-3)/2-j},
\quad \forall z>0,\, 0\le j\le n-3,\eqno{(B.32)}$$
 $$\left|\partial_z^jb_{\nu}^\pm(z)\right|\le C_\epsilon z^{-\epsilon}
\langle z\rangle^{-(n-1)/2+\epsilon},
\quad \forall z>0,\, j=n-2,\eqno{(B.33)}$$
  $$\left|\partial_z^jb_{\nu}^\pm(z)\right|\le C_jz^{n-2-j}
\langle z\rangle^{-(n-1)/2},
\quad \forall z>0,\, j\ge n-1.\eqno{(B.34)}$$
Set 
$$A_\pm^{(1)}(\lambda_1,\lambda_2,\sigma_1,\sigma_2)=
\lambda_1e^{\mp i\sigma_1\lambda_1}
\partial_{\lambda_1^2}^k
\partial_{\lambda_2^2}^m\left(\rho(\lambda_1^2,\lambda_2^2)e^{\pm i\sigma_1
\lambda_1}
b^\pm_\nu(\sigma_1\lambda_1){\cal J}_\nu(\sigma_2\lambda_2)\right),$$
$$A_\pm^{(2)}(\lambda_1,\lambda_2,\sigma_1,\sigma_2)=
\lambda_2e^{\mp i\sigma_2\lambda_2}
\partial_{\lambda_1^2}^m
\partial_{\lambda_2^2}^k\left(\rho(\lambda_1^2,\lambda_2^2)
{\cal J}_\nu(\sigma_1\lambda_1)e^{\pm i\sigma_2\lambda_2}
b^\pm_\nu(\sigma_2\lambda_2)\right).$$
By (B.26), (B.30)-(B.34), we have (with $\ell=0,1$)
$$\left|\partial_{\lambda_1}^\ell A_\pm^{(1)}(\lambda_1,\lambda_2,
\sigma_1,\sigma_2)\right|$$ $$
\le C\langle\sigma_1\rangle^{k+(n-3)/2}
\sigma_2^{n-2}\langle\sigma_2\rangle^{m-(n-1)/2}\langle
 \lambda_1\rangle^{(n-3)/2-k-1},\quad\forall(\lambda_1,\lambda_2),
\eqno{(B.35)} $$
$$\left|\partial_{\lambda_2}^\ell A_\pm^{(2)}(\lambda_1,\lambda_2,
\sigma_1,\sigma_2)\right|$$ $$
\le C\sigma_1^{n-2}\langle\sigma_1\rangle^{m-(n-1)/2}
\langle\sigma_2\rangle^{k+(n-3)/2}\langle
 \lambda_1\rangle^{(n-3)/2-m-2},\quad\forall(\lambda_1,\lambda_2).
\eqno{(B.36)}$$
Using the inequality
$$\|\widehat\varphi(\tau)\|_{L^1}\le C\|\langle\tau\rangle
\widehat\varphi(\tau)\|_{L^2}\le C\sum_{\ell=0}^1\left\|\partial_\lambda^\ell
\varphi(\lambda)\right\|_{L^2}\le C\sum_{\ell=0}^1\sup_{\lambda}\,
\langle\lambda\rangle\left|\partial_\lambda^\ell
\varphi(\lambda)\right|,$$
we obtain from (B.28) and (B.35) (if $k>(n-3)/2$)
$$\left|W_1^{(1)}(\sigma_1,\sigma_2,t,\gamma)\right|\le \sum_\pm Ct^{-k-1/2}
\gamma^{-m}(\sigma_1\sigma_2)^{-2\nu}$$ $$\times\int_{-\infty}^\infty\left|
\int_0^\infty\int_0^\infty
e^{i\tau\lambda_1+i\gamma\lambda_2^2}A_\pm^{(1)}(\lambda_1,\lambda_2,
\sigma_1,\sigma_2)
d\lambda_1d\lambda_2^2\right|d\tau$$ 
 $$\le \sum_\pm\sum_{\ell =0}^1 Ct^{-k-1/2}
\gamma^{-m}(\sigma_1\sigma_2)^{-2\nu}
\sup_{\lambda_1,\lambda_2}\,
\langle\lambda_1\rangle
\left|\partial_{\lambda_1}^\ell A_\pm^{(1)}(\lambda_1,\lambda_2,
\sigma_1,\sigma_2)\right|$$ 
$$\le Ct^{-k-1/2}
\gamma^{-m}\sigma_1^{-(n-2)/2}\langle\sigma_1\rangle^{k+(n-3)/2}
\langle\sigma_2\rangle^{m-(n-1)/2},\eqno{(B.37)}$$
where we have made a change of variables $\tau\to\tau\pm\sigma_1$. 
Similarly, by (B.29) and (B.36), we get (if $m>(n-3)/2$)
$$\left|W_1^{(2)}(\sigma_1,\sigma_2,t,\gamma)\right|\le \sum_\pm Ct^{-k-1/2}
\gamma^{-m}(\sigma_1\sigma_2)^{-2\nu}$$ $$\times\int_{-\infty}^\infty\left|
\int_0^\infty\int_0^\infty
e^{i\tau\lambda_2+i\gamma\lambda_1^2}A_\pm^{(2)}(\lambda_1,\lambda_2,
\sigma_1,\sigma_2)
d\lambda_2d\lambda_1^2\right|d\tau$$ 
 $$\le \sum_\pm\sum_{\ell =0}^1 Ct^{-k-1/2}
\gamma^{-m}(\sigma_1\sigma_2)^{-2\nu}
\sup_{\lambda_2}\,\int_0^\infty
\left|\partial_{\lambda_2}^\ell A_\pm^{(2)}(\lambda_1,\lambda_2,
\sigma_1,\sigma_2)\right|
d\lambda_1^2$$ 
$$\le Ct^{-k-1/2}
\gamma^{-m}\sigma_2^{-(n-2)/2}\langle\sigma_2\rangle^{k+(n-3)/2}
\langle\sigma_1\rangle^{m-(n-1)/2}.\eqno{(B.38)}$$
We would like to apply (B.37) and (B.38) with 
$k=(n-1)/2$, $m=(n-3)/2+\epsilon$,
$0<\epsilon\ll 1$. To this end, we need to show that these estimates are
valid for all real $(n-3)/2<m\le (n-2)/2$, $(n-2)/2\le k<n/2$
 if $n$ is even, and for 
$k=(n-1)/2$ and all real $(n-3)/2<m\le (n-1)/2$ if $n$ is odd. 
This can be done by interpolation as follows. Let 
$\phi\in C_0^\infty({\bf R})$, $\phi(\lambda)=1$ for $|\lambda|\le 1$,
$\phi(\lambda)=0$ for $|\lambda|\ge 2$. Decompose $W_1^{(j)}$ as 
$X^{(j)}+Y^{(j)}$, $j=1,2$, where $X^{(j)}$ and $Y^{(j)}$ are defined
by replacing in the definition of $W_1^{(j)}$ the function $\rho$
by $\phi(\lambda_j)\rho$ and $(1-\phi)(\lambda_j)\rho$, respectively. 
Clearly, the functions $X^{(j)}$ satisfy (B.37) and (B.38), respectively,
 for all integers $0\le k,m<n/2$, while the functions $Y^{(j)}$ satisfy 
(B.37) and (B.38) for all integers $k>(n-3)/2$, $(n-3)/2<m<n/2$, 
respectively. When $n$ is odd, this is fulfilled with $k=(n-1)/2$. 
To show this in the case of even $n$, we write the function $\phi$
as
$$\phi(\lambda)=\sum_{p=0}^\infty\phi_1(2^p\lambda),$$
with some function $\phi_1\in C_0^\infty({\bf R})$, $\phi_1(\lambda)=0$
in a neighbourhood of $\lambda=0$. Thus, 
$$X^{(j)}=\sum_{p=0}^\infty X^{(j)}_p,$$
where $X^{(j)}_p$ is defined by replacing in the definition of 
$X^{(j)}$ the function $\phi(\lambda_j)$ by $\phi_1(2^p\lambda_j)$.
As above, one can see that the functions $X^{(j)}_p$, $j=1,2$, satisfy
(B.37) and (B.38), respectively, with an extra factor in the RHS of the
form $2^{p(k-n/2)}$ for all integers $k\ge (n-2)/2$, and hence, 
by interpolation, for all real $k\ge (n-2)/2$. Therefore, summing up these
estimates we conclude that $X^{(j)}$, $j=1,2$, satisfy
(B.37) and (B.38), respectively, for all real $(n-2)/2\le k<n/2$, 
and in particular for $k=(n-1)/2$. Hence, so do the functions 
 $W_1^{(j)}$. Furthermore, $W_1^{(1)}$ satisfies (B.37) for all integers
$0\le m<n/2$, and hence, by interpolation, for all real $0\le m\le (n-1)/2$
if $n$ is odd, and for all real $0\le m\le (n-2)/2$ if $n$ is even. 
In particular, this is valid with $m=(n-3)/2+\epsilon$. To show that 
the function $W_1^{(2)}$ satisfies (B.38) with $m=(n-3)/2+\epsilon$, 
we decompose it as $Z+N$, where $Z$ and $N$ are defined by replacing in the
definition of $W_1^{(2)}$ the function $\rho$
by $\phi(\lambda_1)\rho$ and $(1-\phi)(\lambda_1)\rho$, respectively.
Clearly, the function $Z$ satisfies (B.37) for all integers
$0\le m<n/2$, and hence, by interpolation, for all real $0\le m\le (n-1)/2$
if $n$ is odd, and for all real $0\le m\le (n-2)/2$ if $n$ is even.
To deal with the function $N$, we write the function $1-\phi$
as
$$(1-\phi)(\lambda)=\sum_{p=0}^\infty\phi_2(2^{-p}\lambda),$$
with some function $\phi_2\in C_0^\infty({\bf R})$, $\phi_2(\lambda)=0$
in a neighbourhood of $\lambda=0$. Thus, 
$$N=\sum_{p=0}^\infty N_p,$$
where $N_p$ is defined by replacing in the definition of 
$N$ the function $(1-\phi)(\lambda_1)$ by $\phi_2(2^{-p}\lambda_1)$.
Now, the functions $N_p$ satisfy (B.38) with an extra factor in the RHS of the
form $2^{-p(m-(n-3)/2)}$ for all integers $0\le m<n/2$, and hence, 
by interpolation, for all real $0\le m\le (n-1)/2$
if $n$ is odd, and for all real $0\le m\le (n-2)/2$ if $n$ is even. 
Therefore, summing up these
estimates we conclude that $N$ satisfies (B.38) for all real 
$(n-3)/2< m\le (n-1)/2$
if $n$ is odd, and for all real $(n-3)/2< m\le (n-2)/2$ if $n$ is even. 
In particular, this is valid with $m=(n-3)/2+\epsilon$.

By (B.37) and (B.38) with $k=(n-1)/2$, $m=(n-3)/2+\epsilon$, we obtain
$$\left|W_1(\sigma_1,\sigma_2,t,\gamma)\right|$$ $$\le Ct^{-n/2}
\gamma^{-(n-3)/2-\epsilon}\left(\sigma_1^{-n+2}
\langle\sigma_1\rangle^{n-2}
\langle\sigma_2\rangle^{-1+\epsilon}+\sigma_2^{-n+2}
\langle\sigma_2\rangle^{n-2}
\langle\sigma_1\rangle^{-1+\epsilon}\right)$$
 $$\le Ct^{-n/2}
\gamma^{-(n-3)/2-\epsilon}\left(\sigma_1^{-n+2}+
\sigma_2^{-n+2}+
\langle\sigma_1\rangle^{-1+\epsilon}+\langle\sigma_2\rangle^{-1+\epsilon}
\right)$$
$$\le Ct^{-n/2}
\gamma^{-(n-3)/2-\epsilon}\left(\sigma_1^{-n+2}+
\sigma_2^{-n+2}+
\sigma_1^{-1+\epsilon}+\sigma_2^{-1+\epsilon}\right),$$
which is the desired bound.
\eproof

Taking $\gamma=h^{\beta'}$ with a suitably chosen constant $\beta'>0$, we
deduce from (B.4), (B.16)-(B.19) and (B.22),
$$\left\|\Psi(t;h)f-F(t)\psi(h^2G_0)f\right\|_{L^\infty}$$ $$\le
Ch^\beta
t^{-n/2}\|f\|_{L^1}+Ch^\beta\left\|\Psi(t;h)f-F(t)\psi(h^2G_0)f
\right\|_{L^\infty},\quad
\forall f\in L^1,\eqno{(B.39)}$$ with some constant $\beta>0$. Taking
$h$ small enough, we can absorb the second term in the RHS of (B.39),
thus obtaining (B.5).
\eproof

Universit\'e de Nantes,
 D\'epartement de Math\'ematiques, UMR 6629 du CNRS,
 2, rue de la Houssini\`ere, BP 92208, 44332 Nantes Cedex 03, France

e-mail: simon.moulin@math.univ-nantes.fr

e-mail: georgi.vodev@math.univ-nantes.fr

\end{document}